\documentclass[a4paper,showpacs, showkeys,11pt]{article}
\usepackage{graphicx}
\usepackage{default}
\title{Efficiency at optimal work from finite reservoirs: \\
a probabilistic perspective}
\author{Ramandeep S. Johal\footnote{ rsjohal@iisermohali.ac.in}\\
Department of Physical Sciences, \\
Indian Institute of Science Education and Research Mohali,\\
Sector 81, Knowledge City, Manauli P.O.,\\ Ajit Garh-140306, Punjab, India.}
\newcommand{\be}{\begin{equation}}
\newcommand{\bea}{\begin{eqnarray}}
\newcommand{\bc}{\begin{center}}            
\newcommand{\ee}{\end{equation}}
\newcommand{\eea}{\end{eqnarray}}
\newcommand{\ec}{\end{center}}
\newcommand{\baa}{\begin{eqnarray*}}
\newcommand{\eaa}{\end{eqnarray*}}
\begin{document}
\baselineskip 18pt

\maketitle
\begin{abstract}
We revisit the classic thermodynamic problem of maximum work extraction from
two arbitrary sized hot and cold reservoirs, modelled as 
perfect gases. 
Assuming ignorance about the extent to which the process
has advanced, which implies an ignorance about the final temperatures, 
we quantify the prior information about the process and assign a prior distribution
to the unknown temperature(s).
This requires that we also take into account the temperature values 
which are regarded to be unphysical in the standard theory, as they
lead to a contradiction with the physical 
laws. Instead in our formulation, such values appear to be consistent  
with the given prior information and hence are included in the inference.
We derive estimates of the efficiency
at optimal work from the expected values of the final temperatures,
and show that these values match with the exact expressions in the limit when 
any one of the reservoirs is very large compared to the other.
For other relative sizes of the reservoirs, we suggest 
a weighting procedure over 
the estimates from two valid inference procedures, that generalizes
the procedure suggested earlier in [J. Phys. A: Math.
Theor. {\bf 46}, 365002 (2013)]. Thus 
 a mean estimate for efficiency is obtained which agrees 
with the optimal performance to a high accuracy.
\end{abstract}
%
\newpage
\section{Introduction}
Maximum work extraction  is a well-known 
problem in classical thermodynamics 
\cite{Thomson, Ondrechen1981, Callenbook,Leff1987b, Vos1992, Lavenda2007}.
It is known that an entropy preserving process yields the upper bound for work.
In recent years, the field of finite-time thermodynamics has 
been intensely investigated, where the power output per cycle is
often sought to be maximum \cite{Broeck2005, Andresen}.
Related to these considerations, the efficiency of engines
at maximum work or power has caught  attention,
in particular, the discussion about its universality
near equilibrium.

In this paper, we consider the issue of efficiency at maximum work
\cite{Leff1987b,Landsberg} from an entirely different, 
probabilistic standpoint. Rather than
performing an optimization of the extracted work, our approach 
follows inductive reasoning or inference 
\cite{Jeffreys1931,Polya1954,Cox}.
The latter has served as a powerful
tool in situations with incomplete information
and is increasingly being applied to the analysis of a wide range of phenomena,
such as in particle physics \cite{Trotta2008}, 
cosmology \cite{Loredo1990}, artificial intelligence 
\cite{Solomonoff1964, Hutter2011} and so on. 

The approach is based on the subjective or 
Bayesian viewpoint \cite{Bernoulli2006,Bayes1763,Laplace1774}
according to which probabilities denote the degree
of rational belief or the state 
of knowledge of an agent. The knowledge which is
available before any data is gathered, is called
the prior information and the degree of belief
about the possible values taken by a parameter
is encapsulated in a prior probability function \cite{Jeffreys1961}.
The basic idea of estimating from prior information
was first proposed in \cite{Johal2010}, in the context of 
quantum thermodynamic machines and later extended
to treat uncertainty in other thermodynamic processes
in Refs. \cite{PRD2012, PRD2013, JRM2013, PRD2014}. A remarkable 
result of these studies is that even with the treatment of
uncertainty in a subjective sense,
the analysis affords  reliable estimates of quantities such as
maximum work as well as the efficiency at maximum work.

In view of the correspondence achieved between the optimal results
and the inference based approach, it seems of importance to
extend the approach to more general situations. 
In this paper, a generalization of Refs. \cite{PRD2013, PRD2014}
is presented which considers the similar problem but
with arbitrary sizes of reservoirs. Recalling the approach,
the central issue was the assignment of the prior
in a constrained thermodynamic process, 
for the uncertain variable such as the temperature.
For the case of identical reservoirs which differ only
in their temperatures, we treated the invariance of the 
prior as the basis for the assignment. The extension
presented in this paper assigns priors by taking 
into account the differences in the reservoirs,
prescribed in the prior information. 
For simplicity of analysis, we illlustrate 
the approach using only the perfect gas model
for the reservoirs.
\section{Efficiency at optimal work}
Consider two perfect gas reservoirs of constant heat capacities
$C_1$ and $C_2$. Let they be at initial temperatures $T_+$ and
$T_- < T_+$. The maximum work is extracted
by removing, in a quasi-static manner,  
a small amount of heat from the hot reservoir,
 converting it to work
with the maximal efficiency, while discarding the waste
heat to the cold reservoir. Thus in a sequence of 
infinitesimal cycles, 
the temperatures of the two reservoirs slowly
approach each other. The process terminates and is said to
be {\it optimal} when the reservoirs achieve  a common 
temperature.

We consider an arbitrary intermediate state
of this process, when the temperatures are
$T_i'$, $i=1,2$. The work extracted upto this stage is:
%
\be
W = C_1 T_+ + C_2 T_- - C_1 T_1' - C_2 T_2',
\label{w}
\ee
where $W > 0$. 
For convenience, we define $C_2/C_1 = x$, $T_-/T_+ = \theta$, 
$T_1 = T_1'/T_+$ and $T_2 = T_2'/T_+$.
So the work is rewritten as
$
W = C_1 T_+ (1+ x \theta - T_1 - x T_2).
$
The constraint of entropy conservation for perfect gases:
$
\Delta S = C_1 \ln T_1 + C_2 \ln T_2 -  C_1 \ln T_+ - C_2 \ln T_- = 0,
$
yields the following one-to-one relation between the two scaled temperatures: 
\be
T_1 = \theta^x {T_2}^{-x}.
\label{t12}
\ee
For the optimal process, the final common temperature is $T_c = \theta^{x/1+x}$.
The {\it efficiency at optimal work} (distinguished by a cap) is:
\be
{\hat\eta}_{\rm op} =  1 - \frac{x(\theta^{x/1+x}-\theta)}{(1-\theta^{x/1+x})}.
\label{eop}
\ee
It is also of interest to consider  efficiency for intermediate
stages.
First, we can use Eq. (\ref{t12}) to express $W$ as a function of
one variable only, such as
\be
W(T_2) = C_1 T_+ \left( 1 + x \theta - \theta^x {T_2}^{-x} - x T_2 \right),
\label{w2}
\ee
where $\{  C_1, T_+, x, \theta \}$ all are fixed for the given process.
One can distinguish two regimes:

a) {\bf  $x<1$}:
In this case, it is convenient to express efficiency in terms of the variable $T_2$.
Thus the heat going ``into'' and the heat going ``out'' of the engine is given as
\bea
Q_{\rm in } &=& C_1 T_+ (1 - T_1), \\
            &=& C_1 T_+ \left( 1 - \theta^x {T_2}^{-x} \right), \\
\label{}
Q_{\rm out} &=& C_2 T_+ (T_2 - \theta). 
\label{}
\eea
The efficiency defined by $\eta_x = 1-{Q_{\rm out } }/{Q_{\rm in}}$,
is given as
\be
\eta_x = 1 -  \frac{x(T_2 -\theta)}{\left( 1 - \theta^x {T_2}^{-x} \right)}.    
\label{exl1}
\ee
In particular, in the limit $x \to 0$, i.e. when the hot reservoir is 
very large compared to the cold one, the efficiency becomes 
\be
\eta_0 = 1 +  \frac{T_2 -\theta}{\ln \left( {\theta}/{T_2} \right)}.
\label{eta0}
\ee
Also in this limit, the temperature of the hot reservoir
is assumed to stay constant at $T_1'= T_+$, or $T_1=1$.
Then for the optimal work extraction, the temperature of the cold reservoir
must  approach this value. 
Thus substituting $T_2 = 1$ in Eq. (\ref{eta0}), we obtain 
\be
{\hat\eta}_0 = 1 +  \frac{(1 -\theta)}{\ln \theta}.
\label{e0}
\ee

b) {\bf $x>1$}:
For this case, it is convenient to use $T_1$ as the variable.
From Eqs. (\ref{t12}) and (\ref{w2}), we have
\be
W(T_1) = C_1 T_+ (1+ x \theta - T_1 - x \theta  {T_1}^{-1/x}),
\label{w1}
\ee
with the efficiency rewrtitten as
\be
\eta_x  = 1 -  \frac{x \theta  ({T_1}^{-1/x} -1)}{\left( 1 - T_1 \right)}.    
\label{et1x}
\ee
In the limit $x \to \infty$,
\be
\eta_{\infty} = 1 +  \frac{\theta \ln T_1 }{ (1-T_1)}.  
\label{einf}
\ee
 This applies to a very large cold reservoir in comparison to a finite hot reservoir.
 Here the temperature of the cold reservoir does not change
 i.e. remains at $T_-$.
 For the optimal process, the temperature of the other reservoir
 must approach this value. 
So substituting $T_1' = T_-$ or $T_1 = \theta$ in Eq. (\ref{einf}),
we obtain the efficiency
\be
{\hat\eta}_\infty = 1 +  \frac{\theta \ln \theta}{(1 -\theta)}.
\label{einff}
\ee 
\begin{figure}[h]
\includegraphics[width=8cm]{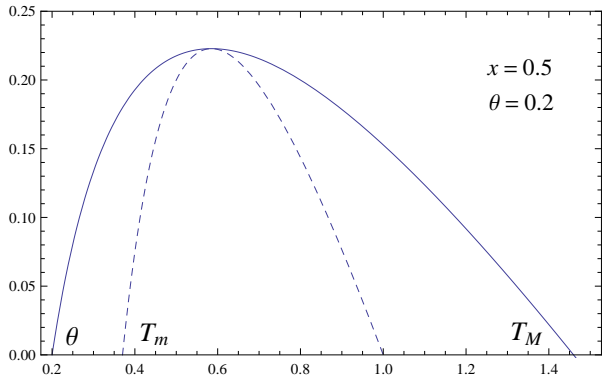}
\caption{Work expressions, Eqs. (\ref{w2}) and (\ref{w1}),
plotted versus the relevant temperature, as solid and dashed
curves respectively, for $W\ge 0$. For $x<1$, the interval  
$[T_m,1]$ of the allowed $T_1$ values, is smaller than the 
corresponding interval $[\theta,T_M]$ for $T_2$.}
\label{w05}
\end{figure}
\begin{figure}[h]
\includegraphics[width=8cm]{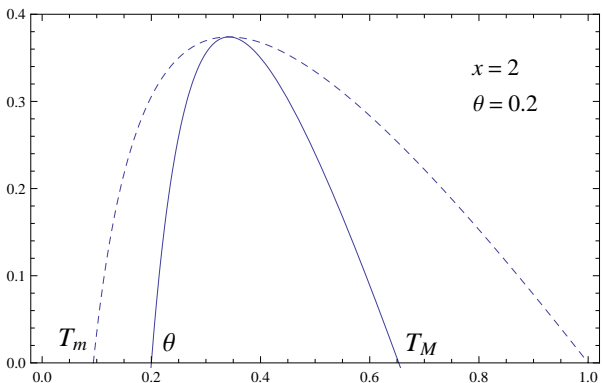}
\caption{ Work expressions plotted versus
temperature, as in Fig. 1, here for $x=2$. For $x>1$, the interval  
of $T_2$ values is smaller than the allowed interval for $T_1$. }
\label{wx2}
\end{figure}
It was observed in \cite{Leff1987b} that the efficiency
at optimal work  is rather insenstive
to the relative sizes of the reservoirs. For most temperature
ranges, it is well approximated by the expression: $1-\sqrt{\theta}$.
We have considered two limiting cases, when one of the reservoirs 
is very large compared to the other. 
Then near equilibrium, i.e. for $\theta$ close to unity,
the efficiency at maximum work,  behaves as $(1-\theta)/2$.
%
\section{Inference}
In this section, we approach the issue of efficiency
at optimal work from the perspective of inference. 
The main approach has been elaborated in Refs. \cite{PRD2013,PRD2014}.
Here our purpose is to seek a generalization for 
different sized reservoirs. 
\subsection{Prior information}
It is instructive to consider the graphical form of Eqs. (\ref{w2}) and (\ref{w1}),
as shown in Figs. 1 and 2 for different $x$ values. In particular, as in Fig. 1
for $x<1$, i.e. for a small cold reservoir and  larger hot reservoir,
the range of $T_1$ values satisfying $W \ge 0$ (dashed curve), is narrower
than the range of $T_2$ values (solid curve).
The two intervals are respectively given as $[T_m, 1]$ and $[\theta, T_M]$.
Due to Eq. (\ref{t12}), we have $T_m = \theta^x {T_M}^{-x}$.
For $x=1$, the two intervals are identical to $[\theta, 1]$.
Further, following Fig. 1, as $x$ approaches zero, the range $[T_m, 1]$
shrinks because $T_m \to 1$, meaning  that as the 
hot reservoir becomes very large, its temperature
tends to remain at its initial value.  Similarly, we can extrapolate 
from Fig. 2 that as $x\to \infty$, $T_M \to \theta$. 

Now imagine a situation where  we are ignorant of the 
final values $T_1$ and $T_2$. We 
adopt the Bayesian approach proposed in \cite{PRD2013, PRD2014}
to deal with this uncertainty. 
In this approach, we first clearly identify the 
 prior information  that we possess on the system:

(i) Whatever the values of $T_1$ and $T_2$, these satisfy
a one-to-one relation, Eq. (\ref{t12}).

ii) Due to condition (i), there is essentially one variable
in the problem, and so an observer is free to formulate 
the problem in terms of either $T_1$ or $T_2$.

(iii) Even when an observer is ignorant of the exact
values of temperatures, his/her state of knowledge 
regarding the true value $T_1$, is the same as the state of an equivalent
observer with respect to $T_2$. 

(iv) An observer possesses knowledge of the 
function $W(T_1)$ or equivalently of $W(T_2)$,
with the condition $W \ge 0$. 
In the physical context, it means 
the set-up works like a heat engine.

The first implication of the condition $W \ge 0$, is that
it fixes the intervals of possible values for the variables
$T_1$ and $T_2$, as shown in Figs. 1 and 2.
It should be emphasized that in our approach,
we also include the temperature values otherwise deemed
unphysical in the standard thermodynamic treatment 
(see Fig. 3). The interval of possible
values is decided solely by the form of the function $W$
alongwith the condition $W \ge 0$. 
\begin{figure}[h]
\includegraphics[width=8cm]{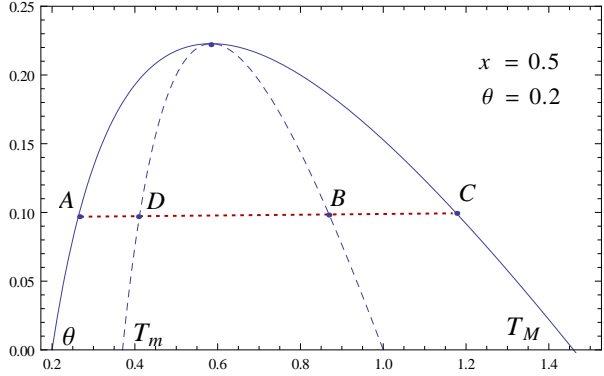}
\caption{Work versus temperature $T_1$ (longdashed curve) and $T_2$ (solid curve). 
For a given value of the work extracted (horizontal dashed line), there are two
possible values each, of $T_2$ (abcissae of points $A$ and $C$) and of $T_1$
(abcissae of $B$ and $D$).
 Only the temperature values corresponding to $A$ and $B$ are
considered physical, because they satisfy the inequality $T_2 < T_1$ and
accord with the natural tendency of heat flow from ``hot to cold''.
The other value of $T_2$ (for the point $C$) and the corresponding value of
$T_1$ (point $D$) do not satisfy this order, and 
usually are disregarded for being unphysical.}
\label{w4sol}
\end{figure}
To close this section, we have considered two observers,
one of whom interprets the uncertainty in terms
of variable $T_2$ while the other, in terms of $T_1$.
Each observer must now quantify the prior information
at his/her command, by assigning  prior
probabilities for the likely values of 
the relevant temperature. 

\subsection{The prior and the estimates}
Further, in view of the points (ii) and (iii) above,
the problem of
the choice of a prior distribution function by either observer
seems to be equivalent.
So for consistency, the same {\it form} of the prior
distribution function would be assigned for each case.
Moreover, the one-to-one relation between
a pair of values $(T_1, T_2)$
suggests that the probability 
of $T_1$ to lie in a small range $[T_1,T_1+dT_1]$
is the same as the probability of $T_2$
to lie in $[T_2,T_2+dT_2]$, where the 
particular values of $T_1$ and $T_2$ are
related by Eq. (\ref{t12}).
Thus we require that
\be
\pi_1( T_m \le T_1 \le 1) dT_1 = \pi_2( \theta \le T_2 \le T_M) dT_2 ,
\ee
where $\pi_i$ is a (normalizable) prior distribution function.
Equivalently, in terms a function $f$, yet to be assigned,
the above condition is
rewritten as
\be
\frac{f(T_1) }{\int_{T_m}^{1} f(T_1) dT_1 }  \left|\frac{dT_1}{dT_2} \right|      = 
\frac{f(T_2) }{\int_{\theta}^{T_M} f(T_2) dT_2 }.
\ee
Finally, the use of Eq. (\ref{t12}) in the above, straightforwardly 
leads to  the form of the prior, by fixing 
$f(T) =1/T$.

Thus for instance, the state of knowledge of observer 2, is expressed by the prior:
\be
\pi_2 (T_2) = \frac{{T_2}^{-1}}{ \ln ({T_M}/{\theta})},
\label{}
\ee
which gives an  expected value of
\be
\overline{T}_2 = \frac{(T_M -\theta)}{\ln ({T_M}/\theta)}.
\label{overt2}
\ee
The estimate for maximum work according to observer 2, is given by 
$W(\overline{T}_2)$ \cite{PRD2013}.
Thereby, the efficiency at maximum work is 
given by replacing $T_2 \to \overline{T}_2$ in Eq. (\ref{exl1}):
\be
{\tilde\eta}_x = 1 -  \frac{x(\overline{T}_2 -\theta)}{\left( 1 - 
\theta^x {\overline{T}_2}^{-x} \right)}.    
\label{exl1es}
\ee
For the behavior of this efficiency near equilibrium, 
see the Appendix.

For general $x$ values, the value $T_M$ is determined by numerically
solving the equation $W(T_2)=0$, Eq. (\ref{w2}), which implies
solving
\be
\frac{\theta}{T_2} = [1 + x(\theta - T_2) ] ^{1/x},
\label{solt2}
\ee
whose trivial solution is $T_2=\theta$.
The other solution is  evaluated numerically.
In the special case of  $x\to 0$, the rhs of Eq. (\ref{solt2})
becomes an exponential function.
In this limit,  $T_M$ is a solution of
\be
T_M \exp (-T_M) = \theta \exp(-\theta).
\label{tmeq}
\ee
It is directly verified that
 Eqs. (\ref{overt2}) and (\ref{tmeq}) together imply that 
the expected value $\overline{T}_2 = 1$.
Thus for $x\to 0$, the average estimate of $T_2$ infers exactly the 
optimal process discussed in Section II, and the estimate for efficiency is the same
as Eq. (\ref{e0}). Again in this limit, the range $[T_m, 1]$
of $T_1$, shrinks to zero and so
 only the inference on temperature $T_2$, needs to be conducted. 

Similarly, the corresponding prior for $T_1$, is
\be
\pi_1 (T_1) = \frac{{T_1}^{-1}}{ \ln (1/{T_m})},
\label{pit1}
\ee
with the expected value 
\be
\overline{T}_1 = \frac{(1-T_m)}{\ln (1/{T_m})}.
\label{overt1}
\ee
Again, the estimate for efficiency at maximum work 
in terms of temperature $T_1$, is to be obtained by 
replacing $T_1$ by $\overline{T}_1$ in Eq. (\ref{et1x}).

For general $x$ values, the value $T_m$ is determined by numerically
solving the equation $W(T_1)=0$. 
For the limit $x\to \infty$, $T_m$ is the solution of
\be
T_m \exp (-T_m/\theta) =  \exp(-1/\theta).
\label{tmeq1}
\ee
whose trivial solution is $T_m = \theta$.
It can be seen that  Eqs. (\ref{overt1}) and (\ref{tmeq1}),
together imply that $\overline{T}_1 = \theta$.
So for this case also, 
we infer an optimal process from the average estimate for $T_1$.
The estimate for efficiency is the same
as $\eta_\infty$ in Eq. (\ref{einf}). Again, in this limit, only the inference
on $T_1$ is to be performed, as the other temperature remains fixed.
\section{Estimates with arbitrary $x$}
We have seen that  when one of the reservoirs
is very large compared to the other, the estimates for efficiency at optimal work
correspond exactly to the values obtained by direct optimization
of work. It is then of interest to see how the estimates of efficiency
compare with the optimal values for arbitrary values of $x$. In Figs. 4 and 5, we
show the estimates made by observer 1 and 2, for 
different $x$ values. Apparently,  
the estimates by the two observers match with the
optimal values when close to equilibrium ($\theta \approx 1$). Further,
for $x<1$, the estimates in terms of $T_2$
lie closer to the optimal value, the agreement being exact 
in the limit $x \to 0$. 
Similarly, for $x>1$, the agreement between the optimal behavior and 
the estimate by observer 1 is better, whereby the agreement 
becomes exact for $x\to \infty$.

\begin{figure}[h]
\includegraphics[width=8cm]{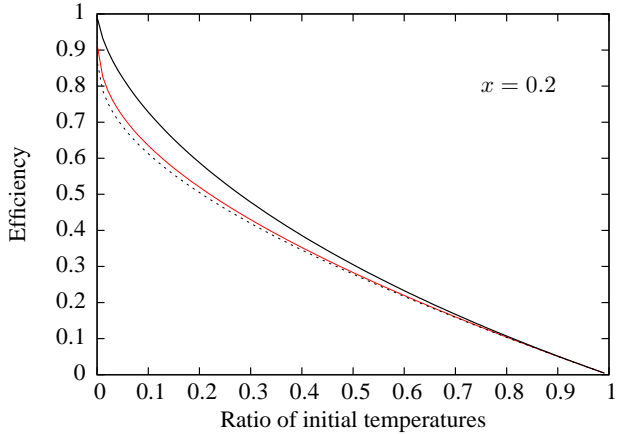}
\caption{Efficiency versus ration of initial temperatures, $\theta$.
Upper curve is the estimated efficiency 
at optimal work due to an observer using $T_1$ as the variable,
the lower curve corresponds to an observer using
$T_2$ variable. The middle curve (red online) is the efficiency
at optimal work, Eq. (\ref{eop}). Note that for $x<1$,
the latter is closer to the estimate
by observer 2.}
\label{eff02}
\end{figure}

\begin{figure}[h]
\includegraphics[width=8cm]{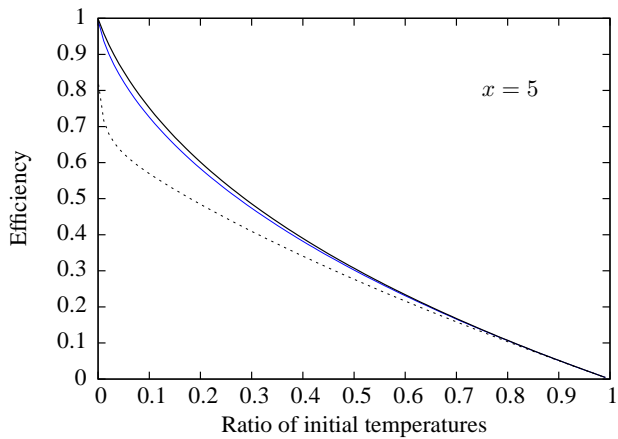}
\caption{ Upper curve is the estimated efficiency 
at optimal work using $T_1$ variable,
the lower curve corresponds to use of
$T_2$ variable. The middle curve (blue online) is the efficiency
at optimal work and is closer to the estimate
of observer 1, when $x>1$.}
\label{eff5}
\end{figure}
\begin{figure}[h]
\includegraphics[width=8cm]{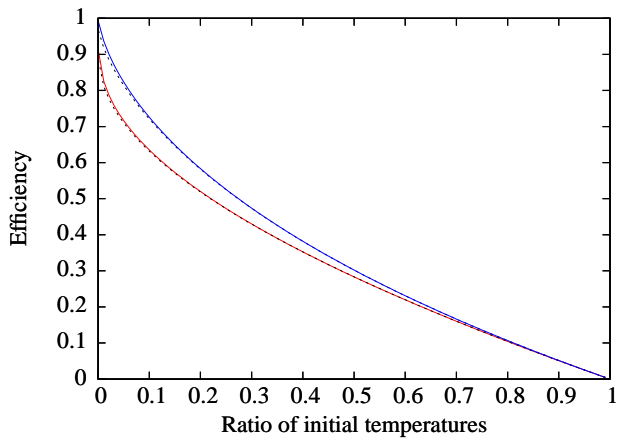}
\caption{The lower curve (red online) for $x=0.2$ and the upper curve 
(blue online) for $x=5$, 
represent efficiency at optimal work as in Figs. 4 and 5.
The dashed curves closely following the above curves are obtained 
by weighing the two estimates of efficiency, as given by Eq. (\ref{etvx}).}
\label{effvth}
\end{figure}

Close to equilibrium, the efficiency at optimal work for arbitrary $x$, 
Eq. (\ref{eop}) behaves as: 
\be
\hat\eta_{\rm op} = \frac{\eta_c}{2} + \frac{(1+ 2 x)}{12(1+x)} \eta_c ^2 +  O[\eta_c ^3] + \cdots
\label{ettax}
\ee
In this regime, the estimates from each observer behave as folllows (see Appendix)
\bea
\tilde\eta_2 &=& \frac{\eta_c}{2} + \frac{\eta_c ^2}{12} + O[{\eta_c ^3}] + \cdots \\
\tilde\eta_1 &=& \frac{\eta_c}{2} + \frac{\eta_c ^2}{6}  + O[{\eta_c ^3}] + \cdots
\eea
These considerations motivate the observation that the 
efficiency at optimal work for arbitrary $x$, can be
estimated as a weighted mean over the estimates due
to each observer.
In fact, let the mean estimate of efficiency at optimal work, be defined as
\be
\tilde{\eta} = \frac{x}{1+x} \tilde\eta_2 +   \frac{1}{1+x} \tilde\eta_1.
\label{etvx}
\ee
This definition is consistent with the exact agreement
observed for the limiting cases discussed in previous sections. 
Further, the agreement with the  efficiency
at optimal work upto second order, Eq. (\ref{ettax}), is also obtained.
The complete expression, Eq. (\ref{eop}) is plotted in Fig. \ref{effvth} and
 compared with the weighted estimate, Eq. (\ref{etvx}). The case of 
 equal sized reservoirs implies equal weights of one-half each,
 which has been discussed in \cite{PRD2013, PRD2014}.

\section{Conclusion}
Motivated by earlier findings, in which the Bayesian
prior probabilities were used to infer the optimal 
performance characteristics of heat engines
and other processes, in this paper we have 
attempted to extend this approach to the scenario
of different-sized reservoirs. Using perfect
gases as the reservoirs, we have reasoned
the appropriate prior which is a renormalised
version of the earlier one derived for 
similar-sized reservoirs. The main focus
of this paper is the estimation of efficiency
at optimal work. It has been shown that the 
estimates of the said
efficiency are exact in the limiting cases
of one reservoir being very large compared
to the other. For intermediate cases, 
a mean estimate for efficiency may be defined,
which reproduces the optimal value close to
equilibrium, correct upto second order of 
the carnot efficiency.

The present generalization further emphasizes the 
relevance of prior information in the success
of the inferential approach. Now the information
on different sizes of the reservoirs, distinguishes 
the two temperatures and assigns each to a 
specific reservoir. In our case, $T_1$
has been assigned to the reservoir with heat capacity
$C_1$ or to the one that was initially hotter ($T_+$).
Note that this information was missing in
earlier studies where similar-sized reservoirs
were considered. 

Further, due to difference in the sizes, the intervals of possible
values for $T_1$ and $T_2$, consistent
with the physical condition of work extraction $W\ge 0$,
are now different. Clearly, the intervals reduce to
the common interval $[\theta,1]$ for similar sizes
of the reservoirs. As discussed in Fig. 3, the inferred interval
of allowed values, includes values which 
are not considered in the physical model. In particular,
the spontaneous flow of heat dictates that the initially
hot reservoir remains hotter than the other, initially
cold reservoir. In this sense, the inference based
analysis is more abstract and does not make use of all the 
physical considerations relevant to the actual
process. This aspect reminds one of
Jaynes' approach to statistical mechanics
where the latter is looked upon as a theory
of statistical inference and physical considerations
of ensembles, reservoirs are not necessary 
for inferring the state of the system
consistent with the given prior information. 
\section*{Appendix}
In the following, we derive the estimates for efficiency
when $\theta$ is close to unity, so that $\eta_c = 1-\theta$
is a small parameter.
For convenience, we take $0<x<1$.
Refering to Fig. 1, 
when the lower bound $\theta$ for $T_2$, is close to unity,
the upper bound $T_M >1$, is also
close to unity. Introducing the small parameter $\omega>0$ as
$T_M / \theta =  1+ \omega$, 
we can rewrite Eq. (\ref{solt2}) as
$1- (1-\eta_c) x \omega = (1+ \omega)^{-x}$.
Applying binomial expansion to the rhs of this equation,  
we obtain upto second order:
\be
(1+x)(2+x) \omega^2 - 3(1+x)\omega + 6 \eta_c = 0,
\ee
whose acceptable solution is 
\be
\omega = \frac{3}{(4+2x)} \left(1-\sqrt{1- \frac{8}{3} \frac{(2+x) \eta_c}{(1+x)}}\right),
\ee
which upto second order in $\eta_c$, can be approximated as:
\be
\omega = \frac{2}{(1+x)} \eta_c + \frac{4}{3} \frac{(2+x)}{(1+x)^2} \eta_c^2.
\label{omsol}
\ee
Then, we can also rewrite $\overline{T}_2$ from Eq. (\ref{overt2}) as
\be
\overline{T}_2  = \frac{\theta \omega}{ \ln (1+\omega)}.
\label{t2om}
              \ee
Now the estimate for efficiency defined as
\be
\tilde{\eta}_2 = 1 -  \frac{x(\overline{T}_2 -\theta)}
{\left( 1 - \theta^x {\overline{T}_2}^{-x} \right)},   
\label{exl1e}
\ee
can be expanded as a series in $\eta_c$, by using Eqs. (\ref{omsol}) and
(\ref{t2om}) in (\ref{exl1e}), to obtain:
\be
\tilde{\eta}_2 = \frac{\eta_c}{2} + \frac{\eta_c ^2}{12} + O[\eta_c^3].
\ee
Note that the initial terms are independent of $x$. 

Similarly, one can show that the estimate for observer 1, derived from 
$\overline{T}_1 
=(1-T_m) / \ln (1/T_m)$, where $T_m = (T_M/\theta)^{-x} = 1 - x \theta \omega$,  behave as
\be
\tilde{\eta}_1 = \frac{\eta_c}{2} + \frac{\eta_c ^2}{6} + O[\eta_c^3].
\ee
\end{document}